\documentstyle[12pt]{article}
\def \qed{\hfill $\vrule height 2.5mm  width 2.5mm depth 0mm $}
\def\beq{\begin{eqnarray}}
\def\eeq{\end{eqnarray}}
\def\nn{\nonumber}
\def\wt{\widetilde}

\def\ld{\lambda}

\def\al{\alpha}

\def\ds{\displaystyle}

\newtheorem{th}{Theorem}

\newtheorem{de}{Definition}

\def\m@th{\mathsurround=0pt}

\def\Fsquare(#1,#2){
\hbox{\vrule$\hskip-0.4pt\vcenter to #1{\normalbaselines\m@th
\hrule\vfil\hbox to #1{\hfill$#2$\hfill}\vfil\hrule}$\hskip-0.4pt
\vrule}}

\def\Addsquare(#1,#2){\hbox{$
	\dimen1=#1 \advance\dimen1 by -0.8pt
	\vcenter to #1{\hrule height0.4pt depth0.0pt\vss%
	\hbox to #1{\hss{%
	\vbox to \dimen1{\vss%
	\hbox to \dimen1{\hss$~#2~$\hss}%
	\vss}\hss}%
	\vrule width0.4pt}\vss%
	\hrule height0.4pt depth0.0pt}$}}

\def\Htwobox(#1,#2){%
	\Fsquare(0.4cm,#1)\Addsquare(0.4cm,#2)}

\def\Hthreebox(#1,#2,#3){%
	\Fsquare(0.4cm,#1)\Addsquare(0.4cm,#2)\Addsquare(0.4cm,#3)}	

\def\Hfourbox(#1,#2,#3,#4){%
    \Hthreebox(#1,#2,#3)\Addsquare(0.4cm,#4)}
    
\def\Hfivebox(#1,#2,#3,#4,#5){%
    \Hfourbox(#1,#2,#3,#4)\Addsquare(0.4cm,#5)}
	
\def\Vfivebox(#1,#2,#3,#4,#5){%
	\normalbaselines\m@th\offinterlineskip
	\vtop{\hbox{\Fsquare(0.4cm,#1)}
	      \vskip-0.4pt
	      \hbox{\Fsquare(0.4cm,#2)}	
	      \vskip-0.4pt
	      \hbox{\Fsquare(0.4cm,#3)}
	      \vskip-0.4pt
	      \hbox{\Fsquare(0.4cm,#4)}		
	      \vskip-0.4pt
	      \hbox{\Fsquare(0.4cm,#5)}}}
	
\def\Vfourbox(#1,#2,#3,#4){%
	\normalbaselines\m@th\offinterlineskip
	\vtop{\hbox{\Fsquare(0.4cm,#1)}
	      \vskip-0.4pt
	      \hbox{\Fsquare(0.4cm,#2)}	
	      \vskip-0.4pt
	      \hbox{\Fsquare(0.4cm,#3)}	
	      \vskip-0.4pt
	      \hbox{\Fsquare(0.4cm,#4)}}}

\def\Vthreebox(#1,#2,#3){%
	\normalbaselines\m@th\offinterlineskip
	\vtop{\hbox{\Fsquare(0.4cm,#1)}
	      \vskip-0.4pt
	      \hbox{\Fsquare(0.4cm,#2)}	
	      \vskip-0.4pt
	      \hbox{\Fsquare(0.4cm,#3)}}}

\def\Vtwobox(#1,#2){%
	\normalbaselines\m@th\offinterlineskip
	\vtop{\hbox{\Fsquare(0.4cm,#1)}
	      \vskip-0.4pt
	      \hbox{\Fsquare(0.4cm,#2)}}}

\def\Twoone(#1,#2,#3){%
	\hbox{
	\normalbaselines\m@th\offinterlineskip
	\vtop{\hbox{\Htwobox({#1},{#2})}
	      \vskip-0.4pt
	      \hbox{\Fsquare(0.4cm,#3)}}}}

\def\Threeone(#1,#2,#3,#4){%
	\normalbaselines\m@th\offinterlineskip
	\vtop{\hbox{\Hthreebox({#1},{#2},{#3})}
	      \vskip-0.4pt
	      \hbox{\Fsquare(0.4cm,#4)}}}
	      
\def\Fourone(#1,#2,#3,#4,#5){%
	\normalbaselines\m@th\offinterlineskip
	\vtop{\hbox{\Hfourbox({#1},{#2},{#3},{#4})}
	      \vskip-0.4pt
	      \hbox{\Fsquare(0.4cm,#5)}}}	      

\def\Threetwo(#1,#2,#3,#4,#5){%
	\normalbaselines\m@th\offinterlineskip
	\vtop{\hbox{\Hthreebox({#1},{#2},{#3})}
	      \vskip-0.4pt
	      \hbox{\Htwobox({#4},{#5})}}}

\def\Twotwo(#1,#2,#3,#4){%
	\normalbaselines\m@th\offinterlineskip
	\vtop{\hbox{\Htwobox({#1},{#2})}
	      \vskip-0.4pt
	      \hbox{\Htwobox({#3},{#4})}}}

\def\Twooneone(#1,#2,#3,#4){%
	\normalbaselines\m@th\offinterlineskip
	\vtop{\hbox{\Htwobox({#1},{#2})}
	      \vskip-0.4pt
	      \hbox{\Fsquare(0.4cm,#3)}
	      \vskip-0.4pt
	      \hbox{\Fsquare(0.4cm,#4)}}}
	      
\def\Twooneoneone(#1,#2,#3,#4,#5){%
	\normalbaselines\m@th\offinterlineskip
	\vtop{\hbox{\Htwobox({#1},{#2})}
	      \vskip-0.4pt
	      \hbox{\Fsquare(0.4cm,#3)}
	      \vskip-0.4pt
	      \hbox{\Fsquare(0.4cm,#4)}
	      \vskip-0.4pt
	      \hbox{\Fsquare(0.4cm,#5)}}}
	      
\def\Twotwoone(#1,#2,#3,#4,#5){%
	\normalbaselines\m@th\offinterlineskip
	\vtop{\hbox{\Htwobox({#1},{#2})}
	      \vskip-0.4pt
	      \hbox{\Htwobox({#3},{#4})}
              \vskip-0.4pt
	      \hbox{\Fsquare(0.4cm,#5)}}}
	      
\def\Threeoneone(#1,#2,#3,#4,#5){%
	\normalbaselines\m@th\offinterlineskip
	\vtop{\hbox{\Hthreebox({#1},{#2},{#3})}
	      \vskip-0.4pt
	      \hbox{\Fsquare(0.4cm,#4)}
              \vskip-0.4pt
	      \hbox{\Fsquare(0.4cm,#5)}}}
\begin{document}
\title{Completeness of Bethe's states for generalized $XXZ$ model, II.}
\author{\\ \Large {Anatol N. Kirillov and Nadejda A. Liskova} \\ \\
{\small {\it Steklov Mathematical Institute,}} \\
{\small {\it Fontanka 27, St.Petersburg, 191011, Russia}}}
\date{ }
\maketitle 
\begin{abstract}
For any rational number $p_0\ge 2$ we prove an identity of 
Rogers-Ramanujan's type. Bijection between the space of states for $XXZ$ 
model and that of $XXX$ model is constructed.
\end{abstract}
\vskip 0.5cm

The main goal of our paper is to study a combinatorial relationship 
between the space of states for generalized $XXZ$ model and that for 
$XXX$ one. In our previous paper [KL] we gave a combinatorial description 
of \hbox{states} for generalized $XXZ$ model in terms of the so-called rigged 
$sl(2)$--$XXZ$ configurations. On the other hand it is well-known that 
when the anisotropy parameter $p_0$ of $XXZ$ model goes to infinity then 
the $XXZ$ model under consideration transforms to the $XXX$ one. We are 
going to describe this transformation from combinatorial point of view in 
the case when $p_0$ is an integer.

A combinatorial completeness of Bethe's states for generalized $XXX$-model 
was proven in [K1] and appears to be a starting point for numerous 
applications to combinatorics of Young tableaux and representation theory 
of symmetric and general linear groups, see e.g. [K2]. Here we mention 
only a "fermionic" formula for the Kostka-Foulkes polynomials, see e.g. 
[K2], and the relationship of the last with $\hat{sl}(2)$- branching 
functions $b_{\ld}^{k\Lambda_0}(q)$, see e.g. [K3]. We will show in 
$\S$1, Theorem 2, that $q$-counting of the number of $XXZ$ states using 
Bethe's ansatz approach [TS], [KR], gives rise to the Rogers-Ramanujan 
type formula for any rational number $p_0>2$.

It seems an interesting problem to find a polynomial version of the 
Rogers-Ramanujan type identity from our Theorem 2.

Another question which we are interested in is to understand a 
combinatorial nature of the limit
$$XXZ ~_{\buildrel\longrightarrow\over{p_0\to +\infty}}~ XXX.
$$

In $\S 2$ we will describe a combinatorial rule which shows how the 
$XXZ$-configurations fall to the $XXX$ pieces. For simplicity we consider 
in our paper only the case $p_0>\sum_ms_m$. General case will be 
considered elsewhere.

\hskip 0.5cm

{\bf \S 1. Rogers-Ramanujan's type identity.}

\vskip 0.5cm

 This paper is a continuation of our previous work [KL]. 
 Let us remind the main definitions, notations and results from [KL].
 
 For fixed $p_0\in{\bf R}$, $p_0\ge 2$ let us define (cf. [TS]) 
 a sequence of real 
 numbers $p_i$ and sequences of integer numbers $\nu_i,m_i, y_i$:
 \begin{eqnarray}
&&p_0:=p_0,~p_1=1,~\nu_i=\left[\frac{p_i}p_{i+1}\right],~
p_{i+1}=p_{i-1}-\nu_{i-1}p_i,~i=1,2,\ldots \\
&&y_{-1}=0,~y_0=1,~~y_1=\nu_0,~y_{i+1}=y_{i-1}+\nu_iy_i,~i=0,1,2,\ldots 
 \\
&&m_0=0,~m_1=\nu_0,~m_{i+1}=m_i+\nu_i,~i=0,1,2,\ldots \\
&& r(j)=i,~\hbox{\rm if}~m_i\le j<m_{i+1},~j=0,1,2,\ldots
 \end{eqnarray}
 It is clear that integer numbers $\nu_i$ define the decomposition of 
 $p_0$ into a continuous fraction
 $$p_0=[\nu_0,\nu_1,\nu_2,\ldots ]=\nu_0+\frac{1}{\nu_1+\ds\frac{1}{\nu_2+
 \ldots}}.
 $$
 Let us define (see Fig. 1) a piecewise linear function $n_j$, $j\ge 0$,
 \begin{equation}
n_j:=y_{i-1}+(j-m_i)y_i,~~\hbox{\rm if}~~m_i\le j<m_{i+1}.
 \end{equation}
 It is clear that for any integer $n>1$ there exists the unique rational 
 number $t$ such that $n=n_t$.
\vskip 0.5cm
 
\setlength{\unitlength}{0.44cm}
\begin{picture}(29,25)(-3.2,-1)
\put(0,0){\vector(0,1){24}}
\put(0,0){\vector(1,0){25}}
\multiput(6,0)(0,1){3}{\line(0,1){0.5}}
\multiput(0,2)(1,0){6}{\line(1,0){0.5}}
\multiput(0,4)(1,0){8}{\line(1,0){0.5}}
\multiput(0,3)(1,0){12}{\line(1,0){0.5}}
\multiput(0,6)(1,0){10}{\line(1,0){0.5}}%
\multiput(0,8)(1,0){21}{\line(1,0){0.5}}
\multiput(0,13)(1,0){17}{\line(1,0){0.5}}
\multiput(0,17)(1,0){19}{\line(1,0){0.5}}
\multiput(0,21)(1,0){26}{\line(1,0){0.5}}
\multiput(8,0)(0,1){4}{\line(0,1){0.5}}%
\multiput(10,0)(0,1){6}{\line(0,1){0.5}}%
\multiput(12,0)(0,1){8}{\line(0,1){0.5}}
\multiput(17,0)(0,1){13}{\line(0,1){0.5}}
\multiput(19,0)(0,1){17}{\line(0,1){0.5}}
\multiput(21,0)(0,1){21}{\line(0,1){0.5}}
\multiput(6.5,3.25)(0.5,0.25){4}{\circle*{0.15}}
\multiput(12.5,8.5)(0.5,0.5){10}{\circle*{0.15}}
\thicklines
\thicklines
\put(0.8,0.4){\vector(2,1){5.2}}
\put(6,2){\vector(1,1){6}}
\put(12,3){\vector(1,2){9}}
\put(21,8){\line(1,4){4}}
\put(-2,2){$y_{i-2}$}
\put(-2,3){$y_{i-1}$}
\put(-4,4){$y_{i-1}\!\! +\!\! y_{i-2}$}
\put(-3.5,6){$y_i\! -\! y_{i-1}$}
\put(-2,8){$y_{i}$}
\put(-3.5,13){$y_i\! +\! y_{i-1}$}
\put(-3.5,17){$y_{i+1}\!\! -\!\! y_i$}
\put(-2,21){$y_{i+1}$}
\put(4.5,-1.2){$m_{i-1}$}
\put(6.7,-1.2){$m_{i-1}\!\! +\!\! 1$}
\put(9.9,-1.2){$m_i\!\! -\!\! 1 $}
\put(12.2,-1.2){$m_{i}$}
\put(15.6,-1.2){$m_i\!\! +\!\! 1$}
\put(21.2,-1.2){$m_{i+1}$}
\put(17.9,-1.2){$m_{i+1}\!\! -\!\! 1$}
\put(25,-1.2){$j$}
\put(-1.5,23.5){$n_j$}
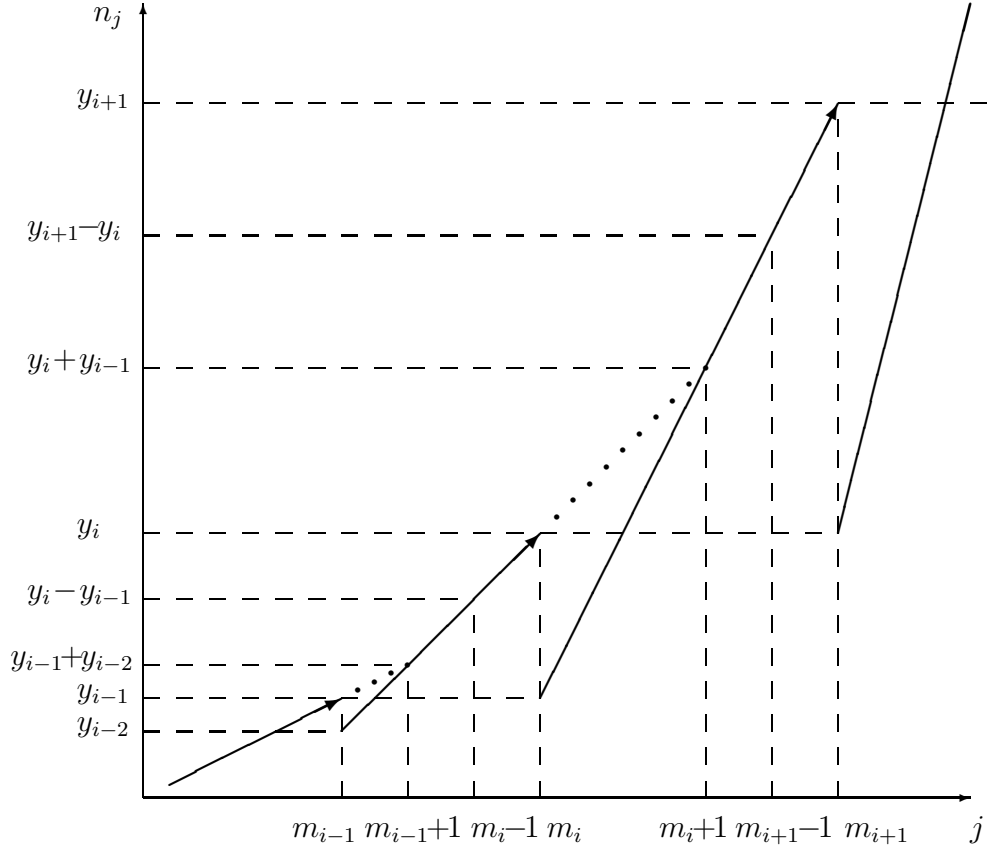
\put(-3,-3.5){Fig.1.~~Image of piecewise linear function $n_j$ in the interval 
$[m_{i-1},m_{i+1}]$}

\end{picture}
\vskip 2cm
 
 Let us introduce additionally the following functions (see [KL])
 \begin{equation}q_j=(-1)^i(p_i-(j-m_i)p_{i+1}),~~\hbox{\rm if}~~m_i
\le j<m_{i+1},
 \end{equation}
 $$\Phi_{k,2s}=\left\{\begin{array}{ll}
 \ds\frac{1}{2p_0}(q_k-q_kn_{\chi}), &\mbox{if $n_k>2s$,}\\
 \ds\frac{1}{2p_0}(q_k-q_{\chi}n_k)+\ds\frac{(-1)^{r(k)-1}}{2}, &\mbox{if 
 $n_k\le 2s$},
 \end{array}\right.
 $$
 where $2s=n_{\chi}-1$.
 
 In order to formulate our main result of part I about the number of Bethe's 
 states for generalized $XXZ$ model, let us consider the following symmetric 
 matrix ~~$\Theta^{-1}=(c_{ij})_{1\le i,j \le m_{\alpha +1}}$:
 \beq
 i)&&c_{ij}=c_{ji}~~\hbox{\rm and}~~c_{ij}=0,
~~\hbox{\rm if}~~|i-j|\ge 2.\nn\\\nn \\
 ii)&&c_{j-1,j}=(-1)^{i-1},~~\hbox{\rm if}~~m_i\le j<m_{i+1}.\nn\\\nn \\
 iii)&&c_{ij}=\left\{\begin{array}{ll}
 2(-1)^i,&\mbox{if $m_i\le j<m_{i+1}-1,~~i\le \alpha$},\\
 (-1)^i,& \mbox{if $j=m_{\alpha +1}$}.\nn
 \end{array}\right.
 \eeq
 \medbreak
 
{\bf Example 1} \ For $p_0=4+\displaystyle{1\over 5}$ 
one can find
\medbreak 
 $$\Theta^{-1}=\left(\begin{array}{ccccccccc}
2 & -1\\ -1 & 2 & -1\\ & -1 & 1 & 1\\ 
&& 1 & -2 &1\\
  &&&1&-2&1\\ &&&&1&-2&1\\ &&&&&1&-2&1\\ &&&&&&1&-1&-1\\ &&&&&&&-1&1
  \end{array}\right)
$$

\bigbreak

 Further, let us consider a matrix $E=(e_{jk})_{1\le j,k\le m_{\alpha 
 +1}}$, where
 $$e_{jk}=(-1)^{r(k)}(\delta_{j,k}-\delta_{j,m_{\al +1}-1}\cdot\delta_{k,m_{\al
+1}}+\delta_{j,m_{\al +1}}\cdot\delta_{k,m_{\al +1}-1}).
$$
Then one can check that (cf. [KL], (3.9))
$$P_j(\ld )+\ld_j=((E-2\Theta ){\wt\ld^t}+b^t)_j,\nn
$$
where $b=(b_1,\ldots ,b_{m_{\al +1}})$ and
$$b_j=(-1)^{r(j)}\left(n_j\left\{{\sum 2s_mN_m-2l\over p_0}\right\}
-\sum_m2\Phi_{j,2s_m}\cdot N_m\right).\nn
$$
 \medbreak
 \begin{th} (\hbox{\rm [KL]}). The number of Bethe's states of generalized $XXZ$ 
 model, $Z^{XXZ}(N,s~|~l)$, is equal to
 \begin{equation}
\sum_{\ld}\prod_j\left(
 \begin{array}{c}
 	((E-B){\wt\ld}^t+b^t)_j  \\ 
 	\ld_j
 \end{array}\right),
 \end{equation}
 where summation is taken over all configurations $\ld =\{\ld_k\}$ such 
 that
 \begin{eqnarray*}
 	 &&\sum_{k=1}^{m_{\al +1}}n_k\ld_k=l,~~\ld_k\ge 0;  \\\nn \\
 	 &&{\wt\ld}=({\wt\ld}_1,\ldots .{\wt\ld}_{m_{\al 
 	 +1}}),~~{\wt\ld}_j=(-1)^{r(j)}\ld_j,~~B=2\Theta.
 \end{eqnarray*}
 \end{th}

One of the main goal of the present paper is to consider a natural 
$q$--analog for (7). Namely, let us define the following $q$--analog of 
the sum (7)
\begin{equation}\sum_{\ld}q^{\frac{1}{2}{\wt\ld}B{\wt\ld}^t}\prod_j\left[
 \begin{array}{c}
 	((E-B){\wt\ld}^t+b^t)_j  \\ 
 	\ld_j
 \end{array}\right]_{q^{\epsilon_j}},
 \end{equation}
where ~~$\epsilon_j=(-1)^{r(j)}$.
 
Let us remind that $\left[
 \begin{array}{c}
 	M \\
 	N
 \end{array}\right]_q$ is the Gaussian $q$--binomial coefficient:
 $$\left[
 \begin{array}{c}
 	M \\
 	N
 \end{array}\right] =\left\{
 \begin{array}{ll}
 	\ds\frac{(M)_q}{(N)_q(M-N)_q}, & \mbox{if $0\le N\le M$},  \\ \\
 	0 & \mbox{otherwise}.
 \end{array}\right.
 $$

{\bf Remark.} In our previous paper [KL], see (5.1) and (5.2), we had 
considered another $q$-analog of (7). It turned out however that the 
$q$-analog (5.1) from [KL], probably, does not possess the good combinatorial 
properties.
 
One of the main results of Part II is the following
\medbreak

\begin{th} (\hbox{\rm Rogers-Ramanujan's type identity}). Assume that 
$p_0$ be a rational number, $p_0\ge 2$,
and
  \begin{equation}
  	V_l(q)=\sum_{\ld}\frac{q^{{\wt\ld}B{\wt\ld}^t}}{\ds\prod_j(q;q^{\epsilon 
  	(j)})_{\ld_j}},
  	\label{8}
  \end{equation}
where summation in (9) is taken over all configurations $\ld 
  =\{\ld_k\}$ such that
  $$l=\sum_{k\ge 1}n_k\ld_k,~~\ld_k\ge 0.
  $$
Then we have
  $$\sum_{k\ge 0}(-1)^kq^{p_0k^2+\frac{k(k-1)}{2}}(1+q^k)=\sum_{l\ge 
  0}q^{\frac{l^2}{p_0}}V_l(q).
  $$
  \end{th}
A proof is a "$q$-version" of that given in [KL], Theorem 4.1.
  
\vskip 0.5cm
  
{\bf \S 2. $XXZ\to XXX$ bijection.}

\vskip 0.5cm

In this section we are going to describe a bijection between the space of 
states for $XXZ$-model and that of $XXX$-model. Let us formulate the 
corresponding combinatorial problem more exactly. First of all as it 
follows from the results of our previous paper, the combinatorial 
completeness of Bethe's states for $XXZ$ model is equivalent to the 
following identity
\begin{equation}
\prod_m(2s_m+1)^{N_m}=\sum_{l=0}^{N}Z^{XXZ}(N,s~|~l),
\end{equation}
where $N=\ds\sum_m2s_mN_m$ and $Z^{XXZ}(N,s~|~l)$ is given by (7). On the 
other hand it follows from the combinatorial completeness of Bethe's 
states for $XXX$ model (see [K1]) that 
\begin{equation}
\prod_m(2s_m+1)^{N_m}=\sum_{l=0}^{\frac{1}{2}N}(N-2l+1)Z^{XXX}(N,s~|~l),
\end{equation}
where $Z^{XXX}(N,s~|~l)$ is the multiplicity of 
$\left(\frac{N}{2}-l\right)$-spin irreducible representation of $sl(2)$ 
in the tensor product
$$V_{s_1}^{\otimes N_1}\otimes\cdots\otimes V_{s_m}^{\otimes N_m}.
$$

Let us remark that both $Z^{XXZ}(N,s~|~l)$ and $Z^{XXX}(N,s~|~l)$ admits 
a combinatorial interpretation in terms of rigged configurations. The 
difference between the space of states of $XXX$ model and that of $XXZ$  
model is the availability of the so-called $1^{-}$-configurations (or 
$1^-$ string) in the space of states for the last model. The presence of 
$1^-$-strings in the space of states for $XXZ$-model is a consequence of 
broken $sl(2)$-symmetry of the $XXZ$-model. Our goal in this section is 
to understand from a combinatorial point of view how the anisotropy of 
$XXZ$ model breaks the symmetry of the $XXX$ chain. More exactly, we 
suppose to describe a bijection between $XXZ$-rigged configurations and 
$XXX$-rigged configurations. Let us start with reminding a definition of 
rigged configurations.
 
We consider at first the case of $sl(2)$ $XXX$-magnet. Given a 
composition $\mu =(\mu_1,\mu_2,\ldots )$ and a natural integer $l$ 
by definition a 
$sl(2)$-configuration of type $(l,\mu )$ is a partition $\nu\vdash l$ 
such that all vacancy numbers 
\begin{equation}
P_n(\nu ;\mu ):=\sum_k\min (n,\mu_k)-2\sum_{k\le n}\nu'_k
\end{equation}
are nonnegative. Here $\nu'$ is the conjugate partition. 
A rigged configuration of type $(l,\mu )$ is a 
configuration $\nu$ of type $(l,\mu )$ together with a collection of 
integer numbers $\{J_{\alpha}\}_{\alpha =1}^{m_n(\nu )}$ such that
$$0\le J_1\le J_2\le\cdots\le J_{m_n(\nu )}\le P_n(\nu ;\mu ).
$$
Here $m_n(\nu )$ is equal to the number of parts equal to $n$ of the 
partition $\nu$. It is clear that total number of rigged configurations 
of type $(l,\mu )$ is equal to
$$Z(l~|~\mu ):=\sum_{\nu\vdash l}\prod_{n\ge 1}\left (
\begin{array}{c}
	P_n(\nu ;\mu )+m_n(\nu )  \\
	m_n(\nu )
\end{array}\right ).
$$ 
The following result was proven in [K1].

\begin{th} Multiplicity of $(N-2l+1)$-dimensional 
irreducible representation of $sl(2)$ in the tensor product 
$$V_{s_1}^{\otimes N_1}\otimes\cdots\otimes V_{s_m}^{\otimes N_m}
$$
is equal to $Z\left(l~|~\underbrace{2s_1,\ldots ,2s_1}_{N_1},\ldots 
,\underbrace{2s_m,\ldots ,2s_m}_{N_m}\right)$.
\end{th} 

{\bf Example 2} \ One can check that
$$V_1^{\otimes 5}=6V_0+15V_1+15V_2+10V_3+4V_4+V_5.
$$ 
In our case we have $\mu =(2^5)$. Let us consider $l=5$. It turns out 
that there exists three configurations 
of type $(3,(2^5))$, namely

\vskip 1cm
\hskip 0.5cm \hbox{
\hbox{\Fsquare(0.4cm,{})\Addsquare(0.4cm,{})\Addsquare(0.4cm,{})\Addsquare(0.4cm,{})\Addsquare(0.4cm,{})~0} \hskip 2cm 
\normalbaselines\m@th\offinterlineskip
	\vtop{\hbox{\Hfourbox({},{},{},{})~0}
	      \vskip-0.4pt
	      \hbox{\Fsquare(0.4cm,{})~1}} \hskip 2cm	      
\normalbaselines\m@th\offinterlineskip
	\vtop{\hbox{\Hthreebox({},{},{})~0}
	      \vskip-0.4pt
	      \hbox{\Htwobox({},{})~2}}
 }
\vskip 1cm

Hence $Z(3~|~(2^5))=1+2+3=6={\rm Mult}_{V_0}\left( V_1^{\otimes 5}\right)$.

Now let us give a definition of $sl(2)$-$XXZ$ configuration. We 
consider in our paper only the case when the anisotropy parameter $p_0$ 
is an integer, $p_0\in{\bf Z}_{\ge 2}$. Under this assumption the 
formulae (5) and (6) take the following form:
\begin{eqnarray*}
&&n_j=j, \ \hbox{\rm if} \ 1\le j<p_0, \ v_j=+1;\\
&&n_{p_0}=1, \ v_{p_0}=-1;\\
&&2\Phi_{k,2s}=\frac{2sk}{p_0}-\min (k,2s), \ \hbox{\rm if} \ 1\le k<p_0, 
\ 2s+1<p_0;\\
&&2\Phi_{p_0,2s}=\frac{2s}{p_0}, \ \hbox{\rm if} \ 2s+1<p_0;\\
&&b_{kj}=k-j, \ \hbox{\rm if} \ 1\le j\le k<p_0;\\
&&b_{kp_0}=1, \ \hbox{\rm if} \ 1\le k<p_0;\\
&&a_j:=a_j(l~|~\mu )=\sum_m\min 
(j,\mu_m)-2l-j\left[\frac{\sum_m\mu_m-2l}{p_0}\right], \ \hbox{\rm if} \ 1\le 
j<p_0;\\
&&a_{p_0}(l~|~\mu )=\left[\frac{\sum_m\mu_m-2l}{p_0}\right].
\end{eqnarray*}
   
\begin{de} A $sl(2)$-$XXZ$-configuration of type $(l,\mu )$ is a pair 
$(\ld ,\ld_{p_0})$, where $\ld$ is a composition with all parts strictly 
less than $p_0$, 
$\ds\sum_{j<p_0}j\ld_j +\ld_{p_0}=l$, and such that all vacancy numbers 
$P_j(\ld ~|~\mu )$ are nonnegative.
\end{de} 

Let us remind ([KL]) that
\begin{eqnarray}
&&P_j(\ld |\mu ):=a_j(l~|~\mu )+2\ds\sum_{j<k<p_0}(k-j)\ld_k+\ld_{p_0}, \ 
\hbox{\rm if} \ j<p_0-1;\\
&&P_{p_0-1}(\ld ~|~\mu ):=a_{p_0-1}(l~|~\mu)+\ld_{p_0};\nonumber \\
&&P_{p_0}(\ld ~|~\mu ):=a_{p_0}(l~|~\mu )+\ld_{p_0-1}.\nonumber 
\end{eqnarray}

{\bf Example 3} \ Let us consider $p_0=6$, $s=\frac{3}{2}$, $N=5$, $l=5$. 
The total number of type $(5,(3^5))$ $sl(2)$--$XXZ$ configurations is 
equal to $12$.

\vskip 1cm

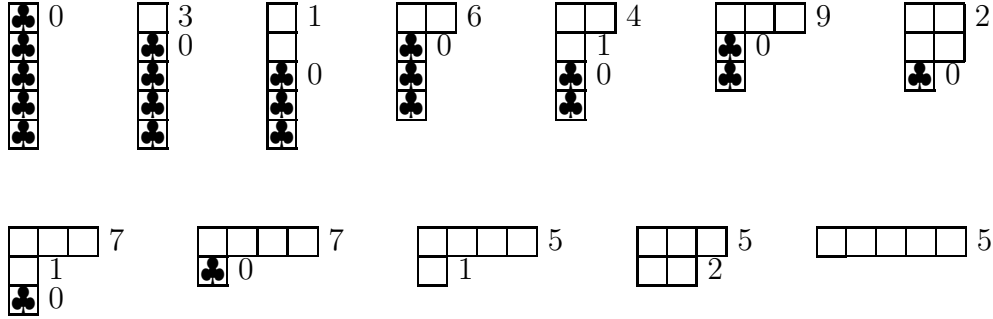
\hbox{
\hbox{\Vfivebox(\hbox{$\clubsuit$},\hbox{$\clubsuit$},\hbox{$\clubsuit$},
\hbox{$\clubsuit$},\hbox{$\clubsuit$})~0}
\hskip 0.7cm
\hbox{\normalbaselines\m@th\offinterlineskip
	\vtop{\hbox{\Fsquare(0.4cm, )~3}
	      \vskip-0.4pt
	      \hbox{\Fsquare(0.4cm,\hbox{$\clubsuit$})~0}	
	      \vskip-0.4pt
	      \hbox{\Fsquare(0.4cm,\hbox{$\clubsuit$})}
	      \vskip-0.4pt
	      \hbox{\Fsquare(0.4cm,\hbox{$\clubsuit$})}		
	      \vskip-0.4pt
	      \hbox{\Fsquare(0.4cm,\hbox{$\clubsuit$})}}}
\hskip 0.7cm
\hbox{\normalbaselines\m@th\offinterlineskip
	\vtop{\hbox{\Fsquare(0.4cm, )~1}
	      \vskip-0.4pt
	      \hbox{\Fsquare(0.4cm, )}	
	      \vskip-0.4pt
	      \hbox{\Fsquare(0.4cm,\hbox{$\clubsuit$})~0}
	      \vskip-0.4pt
	      \hbox{\Fsquare(0.4cm,\hbox{$\clubsuit$})}		
	      \vskip-0.4pt
	      \hbox{\Fsquare(0.4cm,\hbox{$\clubsuit$})}}}
\hskip 0.7cm
\normalbaselines\m@th\offinterlineskip
	\vtop{\hbox{\Htwobox({},{})~6}
	      \vskip-0.4pt
	      \hbox{\Fsquare(0.4cm,\hbox{$\clubsuit$})~0}
	      \vskip-0.4pt
	      \hbox{\Fsquare(0.4cm,\hbox{$\clubsuit$})}
	      \vskip-0.4pt
	      \hbox{\Fsquare(0.4cm,\hbox{$\clubsuit$})}}
\hskip 0.7cm
\normalbaselines\m@th\offinterlineskip
	\vtop{\hbox{\Htwobox({},{})~4}
	      \vskip-0.4pt
	      \hbox{\Fsquare(0.4cm,)~1}
	      \vskip-0.4pt
	      \hbox{\Fsquare(0.4cm,\hbox{$\clubsuit$})~0}
	      \vskip-0.4pt
	      \hbox{\Fsquare(0.4cm,\hbox{$\clubsuit$})}}
\hskip 0.7cm
\normalbaselines\m@th\offinterlineskip
	\vtop{\hbox{\Hthreebox({},{},{})~9}
	      \vskip-0.4pt
	      \hbox{\Fsquare(0.4cm,\hbox{$\clubsuit$})~0}
              \vskip-0.4pt
	      \hbox{\Fsquare(0.4cm,\hbox{$\clubsuit$})}}
\hskip 0.7cm
\normalbaselines\m@th\offinterlineskip
	\vtop{\hbox{\Htwobox({},{})~2}
	      \vskip-0.4pt
	      \hbox{\Htwobox({},{})}
              \vskip-0.4pt
	      \hbox{\Fsquare(0.4cm,\hbox{$\clubsuit$})~0}}
}

\vskip 1cm

\hbox{
\normalbaselines\m@th\offinterlineskip
	\vtop{\hbox{\Hthreebox({},{},{})~7}
	      \vskip-0.4pt
	      \hbox{\Fsquare(0.4cm,)~1}
             \vskip-0.4pt
	      \hbox{\Fsquare(0.4cm,\hbox{$\clubsuit$})~0}}
\hskip 0.7cm
\normalbaselines\m@th\offinterlineskip
	\vtop{\hbox{\Hfourbox({},{},{},{})~7}
	      \vskip-0.4pt
	      \hbox{\Fsquare(0.4cm,\hbox{$\clubsuit$})~0}}
\hskip 0.7cm
\normalbaselines\m@th\offinterlineskip
	\vtop{\hbox{\Hfourbox({},{},{},{})~5}
	      \vskip-0.4pt
	      \hbox{\Fsquare(0.4cm,)~1}}
\hskip 0.7cm
\normalbaselines\m@th\offinterlineskip
	\vtop{\hbox{\Hthreebox({},{},{})~5}
	      \vskip-0.4pt
	      \hbox{\Htwobox({},{})~2}}
\hskip 0.7cm
\Hfivebox( , , , , )~5
}
\vskip 1cm     	      

The total number of type $(5,(3^5))$ rigged configurations is equal to 
$$Z^{XXZ}(5~|~(3^5))=101=1+4+3+7+10+10+6+16+8+12+18+6.
$$
Here we used a symbol \ $\clubsuit$ \ to mark a $1^-$--strings.

Now we are ready to describe a map from the space of states for $XXZ$ 
model to that of $XXX$ one. More exactly we are going to describe a rule 
how a $XXZ$-configuration fall to the $XXX$-pieces. At first we describe this 
rule schematically: 

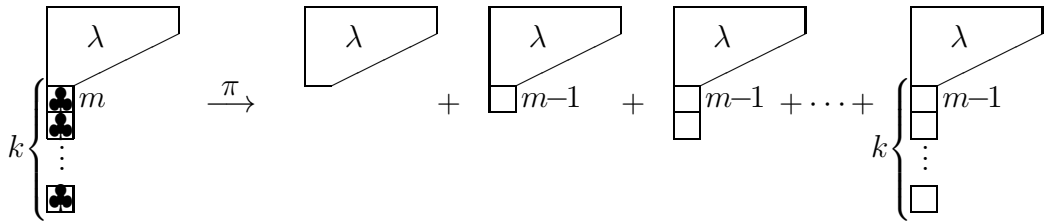
\setlength{\unitlength}{0.35cm}
\begin{picture}(25,12)(1.5,-1.75)
\put(1.2,5){\line(1,0){1}} 
\put(1.2,8){\line(1,0){5}} 
\put(6.2,8){\line(0,-1){1}}
\put(2.2,5){\line(2,1){4}} 
\put(1.2,5){\line(0,1){3}} 
\put(-0.3,2.3){$k\!\left\{\begin{array}{c}~\\~\\~\\~\\ \end{array}\right.$}
\put(1.6,1.8){$\vdots$}
\put(2.4,4.2){$m$}
\put(1.2,3){\line(0,1){2}}
\put(1.2,3){\line(1,0){1}}
\put(2.2,3){\line(0,1){2}}
\put(1.2,4){\line(1,0){1}}
\put(1.25,3.1){$\clubsuit$}
\put(1.25,4.1){$\clubsuit$}
\put(1.25,0.4){$\clubsuit$}
\put(1.2,0.2){\line(1,0){1}}
\put(1.2,0.2){\line(0,1){1}}
\put(1.2,1.2){\line(1,0){1}}
\put(2.2,0.2){\line(0,1){1}}

\put(11,5){\line(1,0){1}} 
\put(11,8){\line(1,0){5}} 
\put(16,8){\line(0,-1){1}}
\put(12,5){\line(2,1){4}} 
\put(11,5){\line(0,1){3}} 

\put(18,5){\line(1,0){1}} 
\put(18,8){\line(1,0){5}} 
\put(23,8){\line(0,-1){1}}
\put(19,5){\line(2,1){4}} 
\put(18,5){\line(0,1){3}} 
\put(18,4){\line(1,0){1}}
\put(18,4){\line(0,1){1}}
\put(19,4){\line(0,1){1}}
\put(19.2,4.2){$m\!\!-\!\!1$}

\put(25,5){\line(1,0){1}} 
\put(25,8){\line(1,0){5}} 
\put(30,8){\line(0,-1){1}}
\put(26,5){\line(2,1){4}} 
\put(25,5){\line(0,1){3}} 
\put(25,3){\line(0,1){2}}
\put(25,3){\line(1,0){1}}
\put(26,3){\line(0,1){2}}
\put(25,4){\line(1,0){1}}
\put(26.2,4.2){$m\!\!-\!\!1$}

\put(34,5){\line(1,0){1}} 
\put(34,8){\line(1,0){5}} 
\put(39,8){\line(0,-1){1}}
\put(35,5){\line(2,1){4}} 
\put(34,5){\line(0,1){3}} 
\put(34,3){\line(0,1){2}}
\put(34,3){\line(1,0){1}}
\put(35,3){\line(0,1){2}}
\put(34,4){\line(1,0){1}}
\put(34,0.2){\line(1,0){1}}
\put(34,0.2){\line(0,1){1}}
\put(34,1.2){\line(1,0){1}}
\put(35,0.2){\line(0,1){1}}
\put(35.2,4.2){$m\!\!-\!\!1$}
\put(32.5,2.3){$k\!\left\{\begin{array}{c}~\\~\\~\\~\\ \end{array}\right.$}
\put(34.4,1.8){$\vdots$}

\put(2.7,6.5){$\lambda$}
\put(12.5,6.5){$\lambda$}
\put(19.5,6.5){$\lambda$}
\put(26.5,6.5){$\lambda$}
\put(35.5,6.5){$\lambda$}
\put(7.2,4){$\buildrel\displaystyle\pi\over\longrightarrow$}
\put(16,4){+}
\put(23,4){+}
\put(29,4){$+\cdots +$}

\end{picture}

This decomposition corresponds to the well-known identity
$$\left[
\begin{array}{c}
	m+k\\
	k
\end{array}\right]_q=\sum_{j=0}^kq^j\left[\begin{array}{c}
	m+j-1\\
	j
\end{array}\right]_q.
$$

In what follows we will assume that $p_0>\sum_ms_m$.

\begin{th} The map $\pi$ is well-defined and gives rise to a bijection 
between the space of states of $XXZ$-model and that of $XXX$ one.
\end{th}

Proof. Let us start with rewriting the formulae (13) for the 
$XXZ$-vacancy numbers in more convenient form, namely,
\begin{eqnarray}
&&P_j^{XXZ}(\wt\nu~|~\mu )=\sum_m\min (j,2s_m)-2\sum_{k\le 
j}\nu_k'-j\left[\frac{\sum_m2s_m-2l}{p_0}\right], \nonumber \\
&&\hbox{\rm if}~~1\le j<p_0-1;\nonumber \\
&& \nonumber \\
&&P_{p_0-1}^{XXZ}(\wt\nu~|~\mu )=p_0\left\{\frac{\sum_m2s_m-2l}{p_0
}\right\}+\left[\frac{\sum_m2s_m-2l}{p_0}\right] +\ld_{p_0}; \\
&&P_{p_0}^{XXZ}(\wt\nu~|~\mu )=\left[\frac{\sum_m2s_m-2l}{p_0}\right] 
+m_{p_0-1}(\nu ).\nonumber
\end{eqnarray}
Here $\mu =(2s_1,\ldots ,2s_m)$ and $\wt\nu$ is a pair $\wt\nu =(\nu 
,\ld_{p_0})$, where $\nu$ is a partition such that $l(\nu )\le p_0-1$, 
$|\nu |+\ld_{p_0}=l$. Relationship between $\ld$ from Definition~1 and 
$\nu$ is the following
$$m_j(\nu )=\ld_j,~~{\rm i.e.}~~\nu =(1^{\ld_1}2^{\ld_2}\ldots 
(p_0-1)^{\ld_{p_0-1}}).
$$
Now let us consider an integer $l\le\sum_ms_m$ and let $\nu\vdash l$ be a 
$XXX$-configuration. Let $\ld_{p_0}$ be integer such that $2\sum 
s_m-2l-p_0<\ld_{p_0}\le\sum_ms_m-l$ and consider the pair $\wt\nu =(\nu 
,\ld_{p_0})$. It is easy to check that
\begin{eqnarray*}
&&P_j^{XXZ}(\wt\nu~|~\mu )=\sum_m\min (j,2s_m)-2\sum_{k\le 
j}\nu_k'=P_j^{XXX}(\nu~|~\mu )\ge 0,\nonumber \\
&&\hbox{\rm if}~~1\le j<p_0-1; \nonumber \\
&& \nonumber \\
&&P_{p_0-1}^{XXZ}(\wt\nu~|~\mu )=\sum_m2s_m-2l+\ld_{p_0}\ge 0; \\
&&P_{p_0}^{XXZ}(\wt\nu~|~\mu )=\ld_{p_0-1}\ge 0.
\end{eqnarray*}
Thus the pair $\wt\nu =(\nu ,\ld_{p_0})$ is a $XXZ$-configuration.
 
Furthermore it follows from our assumptions (namely, $\sum_ms_m<p_0$,\break 
$\ld_{p_0}>0$) that $\ld_{p_0-1}=0$ and both $1^-$-strings and 
$(p_0-1)$-strings do not give a contribution to the space of 
$XXZ$-states. Thus we see that both $XXX$-configuration $\nu$ and 
$XXZ$-configuration $\wt\nu =(\nu ,\ld_{p_0})$ defines the same number of 
states. Now, if $\wt\nu =(\nu ,\mu )$ is a $XXZ$-configuration then $\nu$ 
is a $XXX$ configuration as well. This is clear because (see (14))
$$P_j^{XXX}(\nu~|~\mu )\ge P_j^{XXZ}(\wt\nu~|~\mu ),~~1\le j\le p_0-1.
$$
By the similar reasons if $(\wt\nu ,\ld_{p_0})$ is a $XXZ$-configuration 
then for any\break $0\le k\le\ld_{p_0}$ the pair $(\wt\nu ,\ld_{p_0}-k)$ is 
also $XXZ$-configuration. It follows from what we say above that $\pi$ is 
the well-defined map. Furthermore there exists one to one correspondence 
between the space of $XXX$-configurations and that of 
$XXZ$-configurations, namely,
$$\nu\leftrightarrow\wt\nu =(\nu ,\ld_{p_0}),
$$
where $\ld_{p_0}=[\sum_ms_m-|\nu |]$.

All others $XXZ$-configurations $(\nu ,k)$ with $0\le k<\sum_ms_m-|\nu |-1$ 
give a contribution to the space of descendants for 
$\nu\leftrightarrow\wt\nu$.

\qed

\vskip 0.5cm

{\bf Acknowledgments.} \ We are pleased to thank for hospitality our 
colleagues from Tokyo University, where this work was completed.

\end{document}